\RequirePackage{fix-cm}
\documentclass[journal]{IEEEtran}
\usepackage[latin9]{inputenc}
\usepackage{amsmath}\usepackage{amssymb}\usepackage{graphicx}
\usepackage{tikz}\usetikzlibrary{automata,arrows,positioning,calc}

\makeatletter

\usepackage{epsfig}\usepackage{amsfonts}\usepackage{array}\usepackage{dcolumn}\usepackage{psfrag}\usepackage{amsfonts}\usepackage{accents}\usepackage{booktabs}
\usepackage{graphicx} 
\usepackage{cite}\usepackage[all]{xy}
\usepackage{dsfont}\usepackage{url}\usepackage{xcolor}\usepackage[nolist]{acronym}
\usepackage[FIGTOPCAP]{subfigure}
\usepackage[footnote,draft,silent,nomargin]{fixme}
\usepackage{algorithm}
\usepackage[noend]{algpseudocode}
\usepackage{amsthm}
\usepackage{mathtools}
\usepackage{bbm}
\usepackage{xfrac}
\usepackage{booktabs}
\usepackage{enumitem}
\usepackage{bm}

\graphicspath{{./figures/}}

\acrodef{SINR}{Signal to Interference and Noise Ratio}
\acrodef{CDF}{Cumulative Distribution Function}
\acrodef{PDF}{Probability Density Function}
\acrodef{uMTC}{ultra-reliable Machine-type Communications}
\acrodef{MTC}{Machine-type Communications}
\acrodef{M2M}{Machine-to-Machine}
\acrodef{CTMC}{Continuous-Time Markov Chain}
\acrodef{RV}{Random Variable}
\acrodef{MTTR}{Mean Time to Restoration}
\acrodef{TTI}{Transmission Time Interval}

\makeatother

\begin{document}

\title{Optimized Interface Diversity for Ultra-Reliable Low Latency Communication (URLLC)}
\author{
\IEEEauthorblockN{
Jimmy~J.~Nielsen\IEEEauthorrefmark{1}, %
Rongkuan~Liu\IEEEauthorrefmark{2}, %
and Petar~Popovski\IEEEauthorrefmark{1}%
}

\IEEEauthorblockA{
\IEEEauthorrefmark{1}APNET section, Department of Electronic Systems, Aalborg University, 9220 Aalborg, Denmark\\
}%
\IEEEauthorblockA{
\IEEEauthorrefmark{2}Communication Research Center, Harbin Institute of Technology, Harbin 150001, China\\
}%
\IEEEauthorblockA{
jjn@es.aau.dk, liurongkuan@hit.edu.cn, petarp@es.aau.dk\\
}%
}



\maketitle
\begin{abstract} 
An important ingredient of the future 5G systems will be Ultra-Reliable Low-Latency Communication (URLLC). A way to offer URLLC without intervention in the baseband/PHY layer design is to use \emph{interface diversity} and integrate multiple communication interfaces, each interface based on a different technology. Our approach is to use coding to seamlessly distribute coded payload and redundancy data across multiple available communication interfaces. We formulate an optimization problem to find the payload allocation weights that maximize the reliability at specific target latency values. By considering different scenarios, we find that optimized strategies can significantly outperform $k$-out-of-$n$ strategies, where the latter do not account for the characteristics of the different interfaces. Our approach is supported by experimental results.
\end{abstract}



\section{Introduction}

A key feature of the upcoming 5G technology is the support for Ultra-Reliable and Low Latency Communication (URLLC) \cite{carvalho2016random}. URLLC may be supported both through the 5G new air interface \cite{ji2017introduction} or through the integration of different existing communication technologies \cite{andrews2014will} \cite{monserrat2015metis}. URLLC will enable the support of new use cases with required packet delivery success probability as high as 5-nines ($1\!-\!10^{-5}$) to 9-nines ($1\!-\!10^{-9}$), while at the same time the acceptable latency may be at the sub-second level or even down to a few milliseconds \cite{ratasuk2015recent}. There are proposals for how to decrease the latency in future cellular systems, e.g., by reducing the \ac{TTI} \cite{lahetkangas2014achieving,tullberg2014towards}, fast uplink access \cite{3GPPTR-36881}, or by puncturing URLLC resources on top of eMBB \cite{ji2017introduction}. While 5G with URLLC support (rel. 16) is still several years from deployment, URLLC can already be achieved through integration of multiple communication technologies. 

The use of multiple communication technologies is conceptually very similar to many existing multipath protocols that increase end-to-end reliability~\cite{qadir2015exploiting}. However, low latency requirements exclude reactive protocols that rely on, e.g. retransmission or backup paths. For low latency, we consider \emph{interface diversity} which is in fact a type of path diversity \cite{apostolopoulos2000reliable}, where each path must use a different communication interface. The closest examples of related work that we have identified are the following. In \cite{yap2012making,yap2013scheduling}, the authors demonstrate the use of Software Defined Networking to distribute application packets across multiple available interfaces to increase application throughput. In \cite{singh2016optimal}, the authors consider fairness optimized multi-link aggregation in heterogeneous wireless systems. Candidate architectures for enabling multi-connectivity and high reliability in 3GPP cellular systems are studied in \cite{michalopoulos2016user} and \cite{ravanshid2016multi}. Most recently, in \cite{wolf2017diversity}, the authors present a physical layer analysis of outage probability in multi-connectivity scenarios.

While the use of multiple interfaces, based on different technologies and potentially using independent paths, clearly improves reliability, we are in this work studying how also latency can be reduced using this technique. If the payload is split in parts and different parts are sent over each interface, it is possible to trade-off latency and reliability according to the targeted application. We demonstrated this principle very simply in previous work \cite{nielsen2016latency} and for the present paper we explore the principle in more details.
Specifically, we extend our previous analyses as follows: 
1) we demonstrate how coding can be exploited to enable flexible splitting of payload across interfaces; 
2) we focus the analysis on $N$ independent wireless interfaces, whereas the previous work focused on a specific scenario with only two wireless interfaces; 
3) we formulate the optimization problem of the optimal payload splitting problem as well as the generic evaluation method and present corresponding numerical results; and 
4) we provide an analytic solution for the optimal split of data between two interfaces that minimizes the expected latency.
5) Finally, we use experimental latency data to validate the proposed methodology.

We initially present the system model and transmission strategies in sec.~\ref{sec:system_model}. The methodology for calculating reliability of the considered strategies is presented together with the optimization problem in sec. \ref{sec:reliability_miftx}. In the following sec. \ref{sec:analysis} we provide an analytical solution to the sub problem of splitting between two interfaces. Numerical results are given and discussed in sec. \ref{sec:results}, after which an experimental validation is presented in sec. \ref{sec:exp_results}. Conclusions are given in sec. \ref{sec:conclusion}.

\section{System model}\label{sec:system_model}
\begin{figure}[htb]
	\centering
	\includegraphics[width=\linewidth]{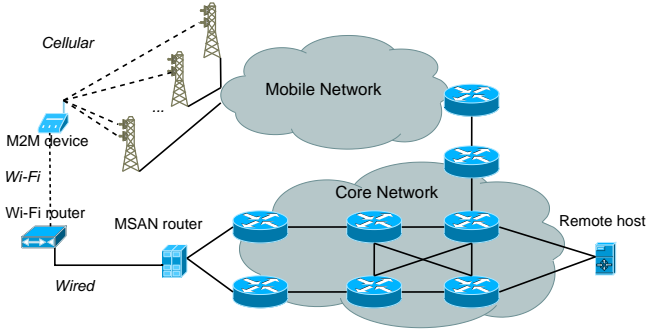}
	\caption{Multiple paths between M2M device (left) and remote host (right).}
	\label{fig:network_diagram}
	\vspace{-9pt}
\end{figure}

We consider a Machine-to-Machine (M2M) device, equipped with $N$ wireless communication interfaces that communicates critical information, e.g. sensor measurements or alarms messages, to a remote host. The model is depicted in Fig. \ref{fig:network_diagram}. In this work we assume that interface failures occur independently and that measurements of end-to-end delay and packet loss are available for the considered interfaces, e.g. through continual network monitoring.

\subsection{Transmission Strategies}\label{sec:strategies}
\begin{figure}[t]
    \centering
    \subfigure[Cloning]{\includegraphics[width=0.4\textwidth]{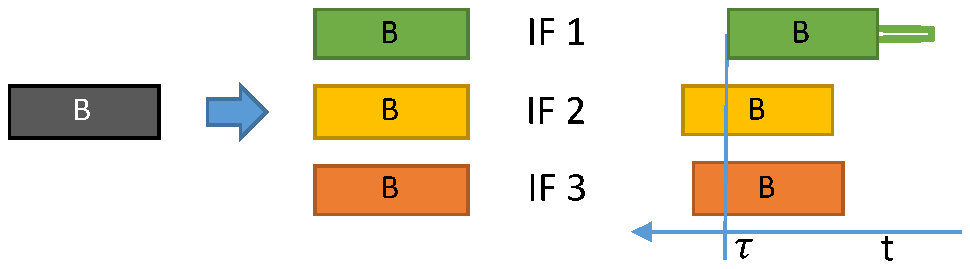}}
    ~ 
    \subfigure[2-out-of-3]{\includegraphics[width=0.4\textwidth]{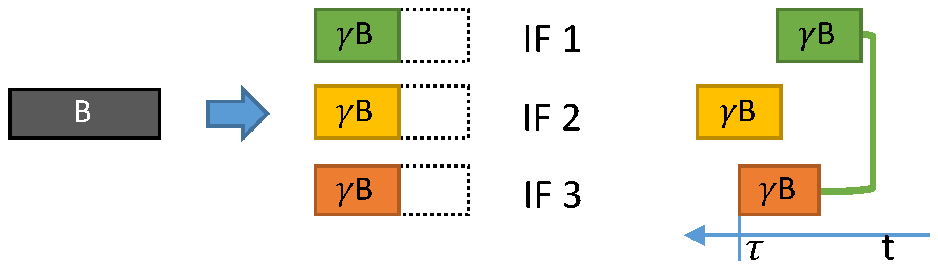}}
    \subfigure[Weighted]{\includegraphics[width=0.4\textwidth]{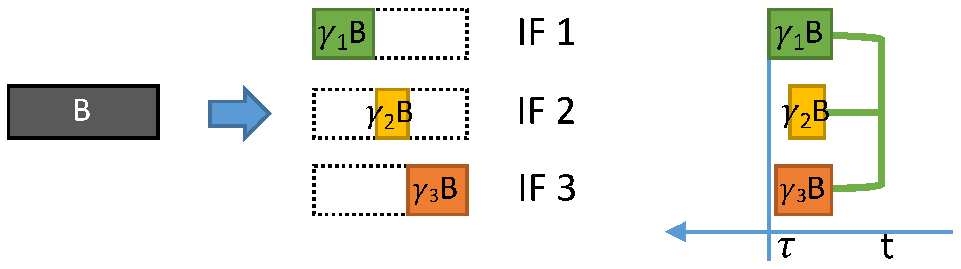}}
    \caption{Transmission strategies, with 2-out-of-3 as example of $k$-out-of-$N$. The time instant $\tau$ is when the payload can be successfully decoded.}
    \label{fig:strategies}
	\vspace{-9pt}
\end{figure}

For transmitting the stream of messages from M2M device to end-host, we consider the following strategies (see Fig.~\ref{fig:strategies}):
\subsubsection{Cloning} In this simple approach, the source device sends a full copy of each message through each of the $N$ available interfaces. Since only one copy is needed at the receiver to decode the message, cloning makes the communication robust at the expense of $N-$fold redundancy.
\subsubsection{Splitting} Instead of sending a full copy on each interface, only a fraction of the message is sent on each interface with this strategy. This allows to trade-off reliability and latency through the selection of the fraction sizes. 
We assume that the payload is encoded, such that we can generate a desired number of coded fragments to be sent through different interfaces. This can be achieved using for example rateless codes \cite{mackay2005fountain} or Reed Solomon codes \cite{wicker1999reed}.
The receiver will be able to decode the encoded message with very high probability as long as it receives coded fragments corresponding to approximately $100 (1+\epsilon) \%$ of the initial message size. A typical value is $\epsilon=0.05$ \cite{mackay2005fountain} and we denote this threshold as $\gamma_\text{d}=1.05$. The coded fragments of a message that are to be sent over the same interface, are grouped together in a single packet to avoid excess protocol overhead.
We assume that for a specific payload message, we let the used code (e.g. rateless or Reed Solomon based) generate coded fragments of a relatively small size, e.g. 10 bytes. 
When nonuniform, \emph{weighted} splitting is used, the challenge is to determine how many fragments to assign to each interface.
Depending on whether identical or different types of interfaces are used, splitting can be realized through either $\bm{k}$-out-of-$\bm{N}$ splitting or weighted splitting, respectively:
	\begin{description}
		 \item [{$\bm{k}$-out-of-$\bm{N}$}] splitting generates $n$ equally sized coded fragments from the payload and the receiver needs to receive at least $k$ of them in order to  decode the message. This strategy allows to trade off reliability and latency, since large redundancy leads to higher reliability but longer transmission times, whereas small redundancy offers a lower error protection but shorter transmission times.
		\item [{Weighted}] the payload is split across interfaces so that the size of the per-interface packet is optimized according to a specific objective. That objective could be to minimize the expected overall transmission latency or to maximize the reliability for a given latency constraint. The optimal solution is, however not trivial, as our analysis shows.
	\end{description}

\subsection{Latency-reliability Function}
Typically, the duration of a packet transmission is depending on the packet size $B$. As a result, we specify the latency-reliability function of interface $i$ as $F_i(x,B)$. This gives the probability of being able to transmit a data packet of $B$ bytes from a source to a destination via interface $i$ within a latency deadline $x$. 
In other words, the value of $F_i(x,B)$ is the achievable reliability $P(X \leq x)$ for a latency $x$ and payload size $B$.
In the following, let $\gamma_i$ specify the fraction of coded payload assigned to interface $i$, where $\gamma_i=[0,\gamma_\text{d}]$. 
Also, let $P_\text{e}$ refer to the long-term error or packet loss probability of an interface, as defined in references \cite{strom20155g,nielsen2016latency}.

\section{Reliability of interface diversity}\label{sec:reliability_miftx}
This section presents the proposed methodologies for achieving reliability through interface diversity. Generally, we assume that the interfaces fail independently, i.e. that the interfaces do not have common error causes.

\subsection{Evaluating reliability for weight assignment}\label{sec:evaluating_weight}
The general approach to evaluating the latency-reliability function for a specific transmission strategy, is that we consider for each possible outcome (in terms of packet losses) if enough payload has been received to decode the message and then sum up the success probability according to the law of total probability. The steps to do this are explained in the following.

Note that payload assignments where $\sum_{i=1}^N \gamma_i < \gamma_\text{d}$ should be avoided, as in such cases, the coded packets can never be decoded.
For enumeration of all possible events, let $\mathbf{C}$ be a $2^N \times N$ matrix listing all possible outcomes for the $N$ interfaces, where a 0 or 1 denotes the successful or failed reception of a packet from the interface of that column:
\begin{equation}
	\textbf{C} = \begin{bmatrix}
		0 & \cdots & 0\\
		0 & \cdots & 1\\
		\vdots & \vdots & \vdots \\
		1 & \cdots & 1
	\end{bmatrix}.
\end{equation}
The element $c_{h,i}$ in the $h$th row and $i$th column of $\mathbf{C}$, refers to the $i$th interface in the $h$th outcome.

For a specific choice of $\bm{\gamma}$, we use the law of total probability to evaluate the resulting latency-reliability function by summing the probability of all successful events. The successful events are the outcomes where the received coded packets can be decoded. The resulting latency-reliability function is:
\begin{equation}\label{eq:weighted_eval}
	F_\text{weighted}(x,\bm{\gamma},B) = \sum\limits_{h=1}^{2^N} d_h \prod\limits_{i=1}^N G_i(x,\gamma_i B)
\end{equation}
where 
\begin{equation}
		d_h = \left\{  \begin{array}{lr}
		1, & \text{if } \sum_{i=1}^N c_{h,i} \cdot \gamma_i \geq 1\\
		0, & \text{otherwise}
		\end{array}\right.
\end{equation}
ensures that we only include outcomes where at least the minimal number of payload fragments are received that allow to decode the payload. Further, $G_i(x)$ is defined as:
\begin{equation}
		G_i(x,\gamma_i B) = \left\{  \begin{array}{ll}
		F_i(x,\gamma_i B), & \text{if } c_{h,i} = 1\\
		1-F_i(x,\gamma_i B), & \text{if } c_{h,i} = 0 .
		\end{array}\right.
\end{equation}

\subsection{Cloning}
For transmissions using packet cloning over $N$ interfaces that can justifiably be considered independent, e.g. cellular connecting to different eNBs or cellular from different operators, we can either use the method presented above or we can use the easier traditional parallel systems \cite{rausand2004system} method to combine the latency-reliability functions as:
\begin{equation}
	F_\text{$N$-clon}(x,\bm{\gamma}, B) = 1-\prod\limits_{i=1}^N (1-F_i(x,\gamma_i B))\label{eq:f_k_par}.
\end{equation}
In either case $\gamma_i=1$ for $i={1, \ldots , N}$.

\subsection{$k$-out-of-$N$ splitting}
While the $k$-out-of-$N$ splitting strategy is only optimal for the case of identical interfaces, it can in principle be used in any case, but with best results in situations where the properties of the available interfaces are comparable.
Generally, we can evaluate the latency-reliability function using the method in sec. \ref{sec:evaluating_weight}, with $\gamma_i = \sfrac{1}{k}$ for $i={1, \ldots, N}$.
In the special case of $N$ identical interfaces, the resulting latency-reliability function can be calculated as:
\begin{equation}
	F_\text{$k$-of-$N$}(x,\bm{\gamma}B) = \sum\limits_{r=k}^N \binom{N}{r} F(x,\gamma B)^r (1-F(x,\gamma B))^{n-r}
\end{equation}
where $\gamma=\sfrac{1}{k}$ and $F(x,\gamma B)$ is the latency-reliability function that represents the identical interfaces.

\subsection{Weighted splitting}
The challenge of the weighted splitting scheme is to determine how many coded fragments to send on each interface to optimize a given utility function. This problem has $N$ degrees of freedom in the form of the payload allocation vector $\bm{\gamma}=\{\gamma_1, \ldots, \gamma_N\}$. 
Formally, this optimization problem can be phrased in the following way:
\begin{equation}\label{eq:opt_weighted}
\begin{array}{rl}
	\underset{\bm{\gamma}}{\max}  & \sum\limits_{r=1}^R F_\text{weighted}(l_r,\bm{\gamma}) \cdot w_r \\
	\text{s.t.} 		& \gamma_i \leq \gamma_\text{d} \\
						& \sum\limits_{i=1}^N \gamma_i \geq \gamma_\text{d}.\\
\end{array}
\end{equation}
where $F_\text{weighted}(l_r,\bm{\gamma})$ is evaluated using eq. \eqref{eq:weighted_eval} and the vectors $\mathbf{l}=\{ l_1, \ldots, l_R\}$ and $\mathbf{w}=\{ w_1, \ldots, w_R\}$ specify the targeted latency values to be maximized and their corresponding importance, respectively. For example, $\mathbf{l}=\{ 0.2, 0.5\}$ and $\mathbf{w}=\{ 1, 10\}$ would mean that reliability at 0.5 s is 10x more important than reliability at 0.2 s.

Assuming that the optimization is solved using a brute-force search, the search space grows as $\left(\sfrac{1}{\delta_\gamma}\right)^N$, where $\delta_\gamma$ is the step size between $\gamma$-values. In practice, the computational tractability of a brute-force search is therefore limited by the number of interfaces $N$ and choice of step size $\delta_\gamma$.
The problem in eq. \eqref{eq:opt_weighted} does not immediately have an analytical solution, since the payload assignment weights in $\bm{\gamma}$ do not translate linearly into specific reliability values. Specifically, when increasing the $\gamma$ value for an interface and thereby increasing the amount of coded payload, the reliability for a specific latency is going to decrease at some point due to the increasing packet size.
However, at the same time a combination of two or more interfaces' increasing $\gamma$-values can add up to $\gamma_\text{d}$ and thereby improve the overall reliability, even if the reliability of the individual interfaces is decreasing as $\gamma$ goes up. 
This behavior, that the overall reliability decreases before it suddenly jumps up, combined with the fact that the $\gamma$ value should be adjusted for each interface individually, narrows the possibilities for analytical solutions.

Therefore, for the numerical results, we include results from a brute-force search that tries out all combinations of $\gamma$-values on the different interfaces, with a step size that is coarse enough to make the search computationally tractable. While we have not managed to solve the whole optimization problem in eq. \eqref{eq:opt_weighted} analytically, we present in the following section an analytical solution to a subproblem of eq. \eqref{eq:opt_weighted}. specifically, we consider how to optimally split coded payload between two interfaces A and B, so that the latency is minimized. 

\section{Analysis of splitting between two interfaces}\label{sec:analysis}
In the optimization problem, we assume the latency of each interface is represented by two Gaussian random variables $ X_{A} \sim \mathcal{N} (\mu_{A}, \sigma_{A}^{2})$ and $ X_{B} \sim \mathcal{N} (\mu_{B}, \sigma_{B}^{2})$. In the following we assume that $\sigma_{A}$ and $\sigma_{B}$ are constant and independent of $\mu_{A}$ and  $\mu_{B}$.

When splitting the payload between two interfaces, the latency is defined by the time at which the last fragment is received. The expected latency is thus the expectation of $\max (X_{A},X_{B})$, which is also the first moment of the random variable $\max (X_{A},X_{B})$. By using the approximation of the expectation of the maximum of two normal random variables from \cite{clark1961greatest}, we obtain
\begin{align}
L = \mathbb{E}[ \max (X_{A},X_{B}) ] = \mu_{A} \Phi (\eta) + \mu_{B} \Phi (-\eta) + \xi \phi (\eta)
\end{align}
where $\phi(x)\!=\!\frac{1}{\sqrt{2 \pi}} \exp^{ -\frac{x^2}{2} }$, $\Phi (x)\!=\!\int_{-\infty}^{x} \phi (t) \mathrm{d}t $,
$\eta\!=\!\frac{ \mu_{A}-\mu_{B} }{ \xi }$, and $ \xi\!=\!\sqrt{ \sigma_{A}^{2} + \sigma_{B}^{2} } $.

To find the minimum of the expected latency, we differentiate $L$ with respect to $\gamma$:
\scalebox{0.825}{\parbox{.5\linewidth}{%
\begin{align}
\frac{\mathrm{d}L}{\mathrm{d}\gamma} &= \frac{\mathrm{d}\mu_{A}}{\mathrm{d}\gamma} \Phi (\eta) + \mu_{A} \phi (\eta) \frac{\mathrm{d}\eta}{\mathrm{d}\gamma} + \frac{\mathrm{d}\mu_{B}}{\mathrm{d}\gamma} \Phi (-\eta) - \mu_{B} \phi (-\eta) \frac{\mathrm{d}\eta}{\mathrm{d}\gamma} + \xi \phi^{\prime} (\eta) \frac{\mathrm{d}\eta}{\mathrm{d}\gamma} \notag \\
&= \frac{\mathrm{d}\mu_{A}}{\mathrm{d}\gamma} \Phi (\eta) + \frac{\mathrm{d}\mu_{B}}{\mathrm{d}\gamma} \Phi (-\eta) + (\mu_{A} \phi (\eta) - \mu_{B} \phi (-\eta) + \xi \phi^{\prime} (\eta)) \frac{\mathrm{d}\eta}{\mathrm{d}\gamma}\notag.
\end{align}}}
Since $\mu_{A} \phi (\eta) - \mu_{B} \phi (-\eta) + \xi \phi^{\prime} (\eta) = 0$, and by using the definition of $\mu$ from eq. \eqref{eq:latency_B} we obtain:
\begin{equation}
\frac{\mathrm{d}L}{\mathrm{d}\gamma} = \frac{\mathrm{d}\mu_{A}}{\mathrm{d}\gamma} \Phi (\eta) + \frac{\mathrm{d}\mu_{B}}{\mathrm{d}\gamma} \Phi (-\eta) = \frac{\alpha_{A}}{2} \Phi (\eta) - \frac{\alpha_{B}}{2} \Phi (-\eta).
\end{equation}

In order to get the optimal solution, $\frac{\mathrm{d}L}{\mathrm{d}\gamma} = 0$ must hold.
So we have the solution as follows:
\begin{align}
\left\{
\begin{array}{lcc}
\Phi (-\eta) = \frac{\alpha_{A}}{\alpha_{A}+\alpha_{B}}, & \mbox{if} & \eta \geq 0 \notag \\
\Phi (\eta) = \frac{\alpha_{B}}{\alpha_{A}+\alpha_{B}}, & \mbox{if} & \eta < 0 \notag
\end{array}
\right.
\end{align}
which is equivalent to:
\begin{align}\label{eq:analytic_splitting}
\left\{
\begin{array}{lcc}
\gamma= \frac{\alpha_{B} + \beta_{B} - \beta_{A} - 2 \xi \Phi^{-1} (\frac{\alpha_{A}}{\alpha_{A}+\alpha_{B}}) }{ \alpha_{A} + \alpha_{B} }, & \mbox{if} & \mu_{A} \geq \mu_{B} \\
\gamma= \frac{\alpha_{B} + \beta_{B} - \beta_{A} + 2 \xi \Phi^{-1} (\frac{\alpha_{B}}{\alpha_{A}+\alpha_{B}}) }{ \alpha_{A} + \alpha_{B} }, & \mbox{if} & \mu_{A} < \mu_{B}.
\end{array}
\right.
\end{align}

\section{Numerical results}
\label{sec:results} 
For the numerical results we will consider the different scenarios specified in Table \ref{tab:scenarios}.
The considered technologies are using the reliability specifications shown in Table \ref{tab:pl_to_rss_and_reliability}.

\begin{table}[bt]
	\centering
	\caption{Linear regression parameters and reliability values.}
	\label{tab:pl_to_rss_and_reliability}
	\begin{tabular}{lcccccc}
	\toprule
		 			& GPRS & EDGE & UMTS & HSDPA & LTE \\ \cmidrule{2-6}
	 	$\alpha$ 	& 0.70 & 0.46 & 0.43 & 0.35 & 0.0067 \\
		$\beta$  	& 400 & 230 & 200 & 178 & 41 \\
		$P_\text{e}$& 0.984 & 0.983 & 0.982 & 0.981 & 0.980 \\ \bottomrule
	\end{tabular}
	\vspace{-9pt}
\end{table}

\begin{table*}[bt]
	\centering
	\caption{Interface and parameter specifications of scenarios $\mathcal{A}$, $\mathcal{B}$, and $\mathcal{C}$.}
	\label{tab:scenarios}
	\begin{tabular}{ccccccccccc}
	\toprule
			& IF1 	& IF2 	& IF3 	& IF4 	& IF5 	& & $B$ & & $\bm{l}$ 	& $\bm{w}$ 	\\ \cmidrule{2-6} \cmidrule{8-8} \cmidrule{10-11}
$\mathcal{A}$  	& UMTS	& GPRS & - 	& - 	& - 	& & 1500 bytes & & $[0 \ldots 1]$ s		& $[0 \ldots 1]$		\\ 
$\mathcal{B}$	& LTE 	& HSDPA & UMTS 	& EDGE 	& GPRS 	& & 1500 bytes & & $[0.1, 0.4, 0.9^*]$ s 	& $[1, 10, 100^*]$		\\
$\mathcal{C}$	& HSDPA	& HSDPA	& GPRS 	& GPRS 	& GPRS 	& & 1500 bytes & & $[0.5]$ s 		& $[1]$		\\ \bottomrule
	\end{tabular}
	\vspace{-9pt}
\end{table*}

While the distribution of latency measurements is usually long-tailed \cite{borella1997self,jacko2000effect}, we will for simplicity use the normal probability distribution to generate latency distributions in the numerical results. While the used probability distribution of  influences the specific results, the methods and general tendencies presented in this paper do not change. Specifically, we assume that the latency of transmissions of packet size $\gamma B$ through a specific interface/path is Gaussian distributed with mean $\mu$ defined as:
\begin{equation}\label{eq:latency_B}
	\mu = \frac{\alpha \cdot \gamma B + \beta}{2} [ms]
\end{equation}
and due to lack of information about the distribution, we assume $\sigma = \frac{\mu}{10}~[ms]$.
The parameters $\alpha$ and $\beta$ characterize the assumed linear relationship between packet size and delay for an interface. The values of $\alpha$ and $\beta$ are shown in Table \ref{tab:pl_to_rss_and_reliability}. The values are derived from field measurements conducted by Telekom Slovenije within the SUNSEED project \cite{sunseed2014web}.

\begin{figure}[t]
	\centering
	\includegraphics[width=0.9\linewidth]{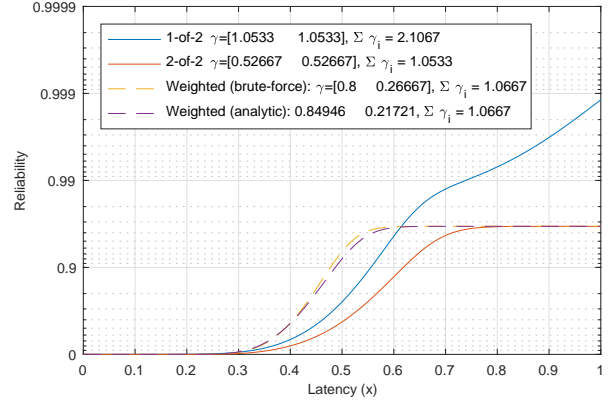}
	\caption{Reliability results for scenario $\mathcal{A}$.}
	\label{fig:scenarioA}
	\vspace{-9pt}
\end{figure}

Initially, we study the simple scenario $\mathcal{A}$, for which we solved the weighted splitting between two interfaces analytically in sec. \ref{sec:analysis}. That is, we used eq. \eqref{eq:analytic_splitting} to determine the optimal splitting threshold $\gamma$. Notice that $\bm{l}$ and $\bm{w}$ are parametrized so that the numerical optimization calculates the expected latency as the analytical optimization.
The results are shown in Fig. \ref{fig:scenarioA}, and show a visually good correspondence between the analytical result and the brute-force search. The brute-force search has a slightly lower expected latency, due to the weight assignment being different. We attribute this minor difference to the use of the approximation of $\mathbb{E}[ \max (X_{A},X_{B}) ]$ from \cite{clark1961greatest}.

In relation to the general idea of splitting, the most important question we seek to answer, is if it makes sense to spend the additional effort required to find the optimal $\gamma$-values for a weighted splitting or if it suffices to use one of the simpler $k$-out-of-$N$ strategies. It is intuitively clear that if the used technologies are all identical, then a $k$-out-of-$N$ strategy will be optimal. But how much better is a weighted scheme in a heterogeneous scenario? To answer this we study three different scenarios that are specified in Table \ref{tab:scenarios}.

\begin{figure}[htb]
	\centering
	\includegraphics[width=\linewidth]{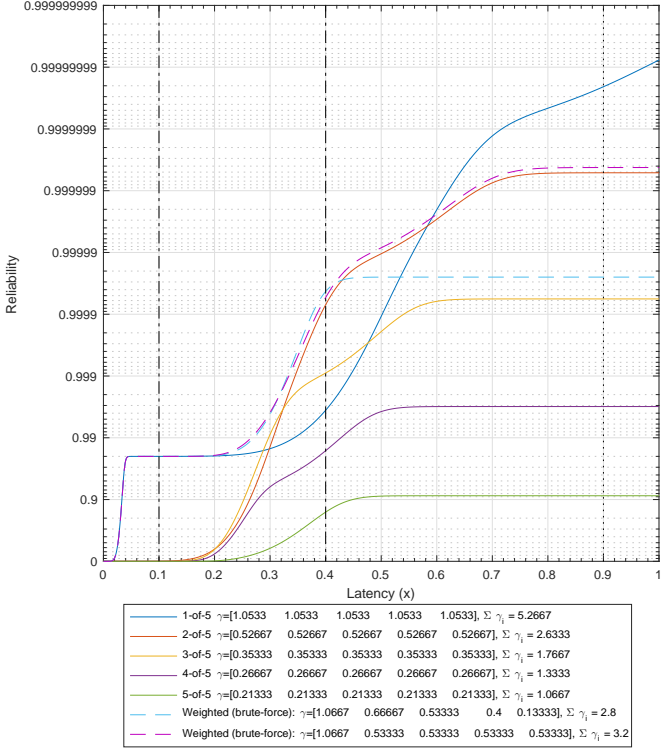}
	\caption{Reliability results for scenario $\mathcal{B}$. Note: the target latency $l_2=0.9$~s only applies to the last strategy.}
	\label{fig:scenarioB}
	\vspace{-9pt}
\end{figure}

The results for scenario $\mathcal{B}$ in Fig. \ref{fig:scenarioB} show two examples of latency-reliability trade-offs that are achieved by considering both when the starred $l$ and $w$ values in Table \ref{tab:scenarios} are included and excluded. In both cases the weighted strategy achieves some reliability in the low latency region ($x<0.2$~s) similar to the 1-out-of-5 strategy and it has the reliability of the 2-out-of-5 strategy around $x=0.4$~s. The difference between the 2 results is that the last one transmits more redundancy data and achieves higher reliability in the $x>0.4$~s region.

\begin{figure}[htb]
	\centering
	\includegraphics[width=\linewidth]{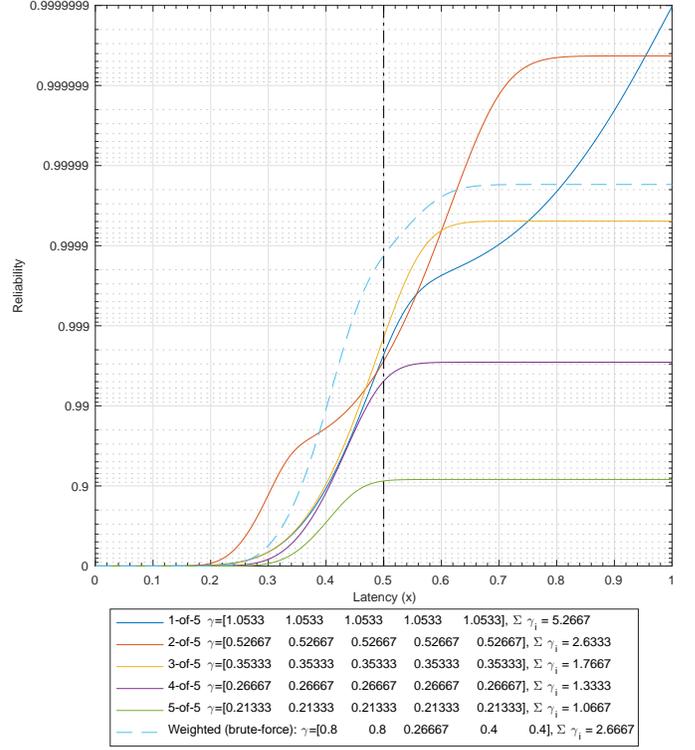}
	\caption{Reliability results for scenario $\mathcal{C}$.}
	\label{fig:scenarioC}
	\vspace{-9pt}
\end{figure}

The results concerning scenario $\mathcal{C}$ that are shown in Fig. \ref{fig:scenarioC} are interesting since they demonstrate a mixed data allocation. This results in the reliability at $x=0.5$~s being 0.9999, which is one decade better than any of the $k$-out-of-$N$ strategies that only go up to 0.999.

\section{Experimental validation}\label{sec:exp_results}
\begin{figure}[htb]
	\centering
	\includegraphics[width=0.9\linewidth]{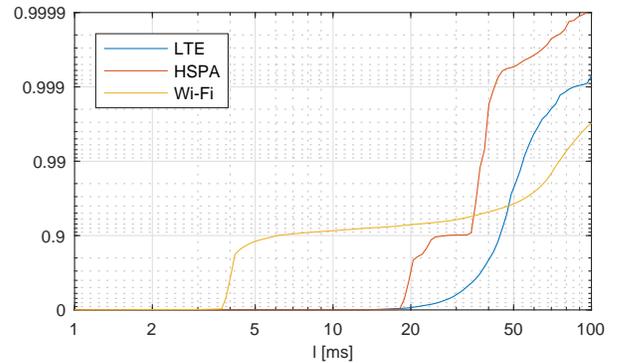}
	\caption{Interfaces' latency-reliability curves. Wi-Fi is IEEE 802.11n.}
	\label{fig:experimental_ifs}
	\vspace{-9pt}
\end{figure}

\begin{figure}[htb]
	\centering
	\includegraphics[width=0.9\linewidth]{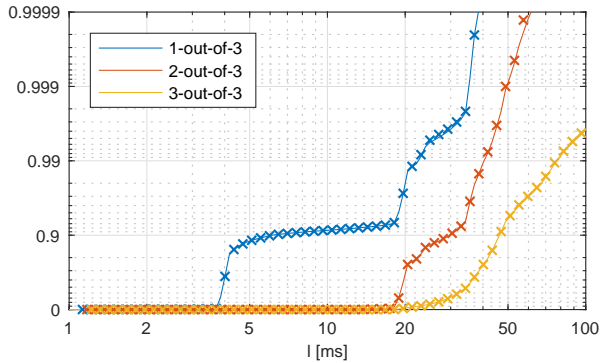}
	\caption{Resulting performance of considered strategies. The lines show the results computed using the method presented in sec. \ref{sec:reliability_miftx}, whereas the crosses show the results of playback-simulation.}
	\label{fig:experimental_lcdfs}
	\vspace{-9pt}
\end{figure}

In addition to the theoretical and model-based results presented above, we have also validated the proposed methods using traces of latency measurements for different communication technologies. Such traces were obtained by sending small (128~bytes) UDP packets every 100~ms between a pair of GPS time-synchronized devices through the considered interface (LTE, HSPA, or Wi-Fi) during the course of a work day at Aalborg University campus. Each trace file can thus be used to playback a time sequence of one-way end-to-end latencies. Our experimental results of multi-interface transmissions are obtained by playing back the three trace files at the same time time in a simulation, where for each 100 ms, the outcome of each considered strategy is recorded. When the playback simulation is done, a latency-reliability curve is calculated for each strategy as the cdf of the recorded outcomes in each 100 ms timestep. This is shown with crosses in Fig. \ref{fig:experimental_lcdfs}. The validation consists in comparing these results to the results that are obtained by using the curves in Fig. \ref{fig:experimental_ifs} to compute the resulting latency-reliability curves using the methods described in sec. \ref{sec:reliability_miftx}. Those results are shown as lines in Fig. \ref{fig:experimental_lcdfs}.

When considering the latency-reliability curves of the interfaces in Fig. \ref{fig:experimental_ifs} it is interesting that HSPA actually performs better than LTE. We believe that this is due to the fact that the majority of current mobile devices connect through LTE if it is available. Thus, the collocated HSPA network experiences a lighter load and allows for quicker access. Another interesting observation is that the Wi-Fi network delivers very low latencies down to below 4~ms for 60\% of packets. However, the 99th percentile latency of 75~ms is higher than both HSPA and LTE. 

From the results in Fig. \ref{fig:experimental_lcdfs}, we see how the 1-out-of-3 strategy is able to outperform any individual interface, as expected.
The plot does not include any result for the Weighted scheme, since the small payload size does not allow for any gain through payload splitting. The lines that represent the theoretical calculation of performance are practically coinciding with the crosses representing the experimental results. This shows that the methods for calculating the resulting performance by relying on the latency-reliability curves of the interfaces, as described in Sec. \ref{sec:reliability_miftx}, indeed produces accurate results when used with actual traffic traces.


\section{Conclusions and Outlook}
\label{sec:conclusion} 
One of the most demanding modes in the upcoming 5G systems will be Ultra-Reliable Low Latency Communication (URLLC). In many cases it should be provided by taking advantage of the fact that multiple communication interfaces are available at the devices. 
In this work we have studied the concept of interface diversity, where multiple communication interfaces and paths are used simultaneously to communicate between two end devices. The use of coding allows us to assign an arbitrary amount of coded payload data to each interface, allowing to trade-off latency and reliability. 
We have formulated the optimization problem to find the payload allocation weights (denoted $\bm{\gamma}$) that maximize the reliability at specific target latency values. We have provided and validated an analytic solution to the subproblem of splitting between two interfaces so that the expected latency is minimized.
By considering different scenarios and numerically solving the full optimization problem for specific target latencies, we have found that optimized strategies can significantly outperform $k$-out-of-$n$ strategies, where the latter do not account for the characteristics of the different interfaces.
Finally, we have experimentally validated the proposed method of computing the resulting performance, and demonstrated the practical gains of interface diversity.



\section*{Acknowledgment}
This work is partially funded by EU, under Grant agreement no. 619437. The SUNSEED
project is a joint undertaking of 9 partner institutions and their contributions are
fully acknowledged. The work was also supported in part by the European Research Council 
(ERC Consolidator Grant no. 648382 WILLOW) within the Horizon 2020 Program.


\bibliographystyle{IEEEtran}
\bibliography{../../Journal/bibliography}


\end{document}